\documentclass[10pt,twocolumn,twoside,english]{IEEEtran}
\usepackage[T1]{fontenc}
\usepackage{verbatim}
\usepackage{amsthm}
\usepackage{amsmath}
\usepackage{amssymb}
\usepackage{esint}
\usepackage{babel}
\usepackage{graphicx}
\usepackage{cite}
\usepackage{caption}

\usepackage{constants}

\makeatletter

\newtheorem{lem}{Lemma}
\newtheorem{thm}{Theorem}

\theoremstyle{plain}

\theoremstyle{remark}
\newtheorem{rem}{Remark}
\theoremstyle{definition}

\usepackage{hyperref}
\def \SU {\operatorname{S}}
\def \A {\operatorname{A}}
\def \B {\operatorname{B}}

\def \Q {\operatorname{Q}}
\def \SR {\operatorname{Y}}
\def \GR {\mathbb{G}}
\def \E {\mathbb{E}}

\def \F {\operatorname{F}} 
\def \wt {\widetilde}

\hypersetup{
   colorlinks=true,%
   citecolor=black,%
   filecolor=black,%
   linkcolor=red,%
   urlcolor=blue
}

\makeatletter
\def\blfootnote{\xdef\@thefnmark{}\@footnotetext}
\makeatother

\def \l{\left}
\def \r{\right}

\begin{document}

\title{What to Expect\\ When You Are Expecting on the Grassmannian}

\author{Armin Eftekhari, Laura Balzano, and Michael B.\ Wakin}

\maketitle

\begin{abstract}
Consider an incoming sequence of vectors, all belonging to an unknown subspace $\SU$, and each with many missing entries. In order to estimate $\SU$, it is common to partition the data into blocks and iteratively update the estimate of $\SU$ with each new incoming measurement block.

In this paper, we investigate a rather basic question: Is it possible to identify $\SU$ by averaging the column span of the partially observed incoming measurement blocks on the Grassmannian?

We show that in general the span of the incoming blocks is in fact a biased estimator of $\SU$ when data suffers from erasures, and we find an upper bound for this bias.
 We reach this conclusion by examining the  defining optimization program for the \emph{Fr\'{e}chet expectation} on the Grassmannian, and with the aid of a sharp  perturbation bound and standard large deviation results.

\end{abstract}

\section{Problem Statement}
\label{sec:problem statement}

Consider an $r$-dimensional subspace $\SU$ with orthobasis $S\in\mathbb{R}^{n\times r}$,  {where $n> r$.}\blfootnote{AE is with the Alan Turing Institute in London. LB is with the Electrical Engineering and Computer Science department at the University of Michigan, Ann Arbor. MBW is with the Electrical Engineering and Computer Science department at Colorado School of Mines. (e-mails: aeftekhari@turing.ac.uk; girasole@umich.edu; mwakin@mines.edu) AE is supported by the Alan Turing Institute under the EPSRC grant EP/N510129/1. LB is supported by ARO Grant W911NF-14-1-0634. MBW is partially supported by NSF grant CCF-1409258 and NSF CAREER grant CCF-1149225. The authors would like to thank Dehui Yang for his involvement in the early phases of this project. }
We wish  to identify  $\SU$ from incomplete data, received sequentially, using only limited memory
 \cite{warmuth2008randomized,arora2012stochastic,yang2015streaming}.
\emph{Streaming subspace identification} from incomplete data finds application in system identification \cite{van2012subspace,liu2009interior} where data commonly suffers from erasures,  {or in monitoring network traffic \cite{lakhina2004diagnosing}, where collecting complete data is infeasible.} Such applications arise also in  imaging, computer vision, and communications   \cite{ardekani1999activation,manolakis2002detection,costeira1998multibody,krim1996two,tong1998multichannel},
 to name a few.

More concretely, for an  integer $T$, let $\{q_t\}_{t=1}^T\subset \mathbb{R}^r$ be independent copies of a random vector $q\in\mathbb{R}^r$.
At time $t\in[1:T]:=\{1,2,\cdots,T\}$,  we observe each entry of  $s_t := S\cdot q_t \in \SU$ with a probability of $p\in(0,1]$, and we collect the measurements in $y_t\in\mathbb{R}^n$, setting the unobserved entries to zero.
To reiterate, our objective is to identify the subspace $\SU$ from the measurement vectors $\{y_t\}_{t=1}^T$. Throughout, we assume that $r=\mbox{dim}(\SU)$ is known \emph{a priori} or estimated from data by other means.

 {The  literature of modern signal processing offers a number of efficient algorithms to solve this problem, including (the new) SNIPE \cite{SNIPE},  GROUSE \cite{balzano2015local}, as well as a generalization of the classic {power method} \cite{mitliagkas2014streaming}. These algorithms partition the incoming measurements into  non-overlapping blocks and iteratively update their estimate of the true subspace $\SU$ with each incoming measurement block.}


 {This paper intends to enhance our understanding of this subject by answering a basic, but hopefully interesting, question about averaging zero-filled data on the Grassmannian manifold of $r$-dimensional subspaces. Estimation on the Grassmannian is a problem of general interest~\cite{hong2017regression}. Meanwhile, zero-filling is a common step with missing data and has been shown to have reasonable estimation properties in some cases~\cite{chatterjee2015matrix} or can be used as algorithmic initialization~\cite{keshavan2010matrix,ganti2015matrix}. 

While a byproduct of our work is an algorithm for subspace identification (that computes the mentioned average), more efficient techniques for subspace identification exist in the literature~\cite{SNIPE,balzano2015local,mitliagkas2014streaming} that utilize the information from previous blocks instead of zero-filling the measurement blocks. We also note that block-based subspace averaging algorithms have been used successfully with fully observed ($p=1$) data~\cite{chakraborty2015recursive,salehian2015efficient,chakraborty2017intrinsic}. With these remarks, let us now state this question in detail.}

For an integer $b\ge r$, suppose we partition the incoming measurements $\{y_t\}_{t=1}^T$ into (non-overlapping) blocks of size $b$, which we denote  by $\{Y_k\}_{k=1}^K\subset\mathbb{R}^{n\times b}$, assuming  that  the number of blocks $K=T/b$ is an integer for simplicity. Each measurement block $Y_k$ is a partially-observed copy of $S_k = S\cdot  Q_k$, where $\{S_k\}_{k=1}^K \subset \mathbb{R}^{n\times b}$ and  the coefficient matrices $\{Q_k\}_{k=1}^K \subset\mathbb{R}^{r\times b}$ are obtained by partitioning $\{s_t\}_{t=1}^T$ and $\{q_t\}_{t=1}^T$ into blocks of size $b$, respectively.

Each measurement block $Y_k$ provides a simple, if not accurate, estimate of the underlying subspace $\SU$. Indeed, let $Y_{k,r}\in\mathbb{R}^{n\times b}$ be a rank-$r$ truncation of $Y_k$, obtained by truncating  all but the largest $r$ singular values of $Y_k$.
 Consider the $r$-dimensional subspace $\SR_k=\mbox{span}(Y_{k,r})$, and recall that $\SR_k$ best approximates $\mbox{span}(Y_k)$ among all $r$-dimensional subspaces. We may  consider $\SR_k$ as an estimate of $\SU$. In particular, $\SR_k=\SU$ when there is no erasure ($p=1$) and $Q_k$ is rank-$r$.

By construction, $\{\SR_k\}_{k=1}^K$ are independent and identically distributed random subspaces on the Grassmannian $\mathbb{G}(n,r)$, the manifold of all $r$-dimensional subspaces of $\mathbb{R}^n$. It is therefore natural to consider the ``average'' of the subspaces $\{\SR_k\}_{k=1}^K$ as an estimate of $\SU$. (As we will see in Section \ref{sec:Averaging}, some care must be taken in defining this average. We will also point out that, under mild conditions,  this average can be updated in a streaming fashion, making the scheme suitable for memory-limited scenarios.)

With this introduction, the present work answers the following question:  {\emph{What is the bias of the expectation of each $\SR_k$ with respect to the true subspace $\SU$?}}
In the next section, we formalize this question and find an upper bound for the bias, with our main result summarized in Theorem \ref{thm:bias} below.

\section{ Expectation on Grassmannian}
\label{sec:Averaging}

Consider the following metric on the Grassmannian \cite{absil2004riemannian}:
If $\{\theta_i(\A,\B)\}_{i=1}^r$ are the principal angles between $r$-dimensional subspaces $\A,\B\in\mathbb{G}(n,r)$,\footnote{Principal angles between subspaces generalize the notion of angle between lines. See  \cite{golub2013matrix} for more details.} their  distance is
\begin{equation}\label{eq:geod dist}
d_{\GR}\l( \A,\B\r) = \sqrt{\frac{1}{r} \sum_{i=1}^r \theta_i\l(\A,\B \r)^2}.
\end{equation}
For example, the  distance between two one-dimensional subspaces
(namely, two lines) is the smaller angle that they make.

We can now define the \emph{Fr\'{e}chet expectation} of $\SR_k$ on $\GR(n,r)$ as the subspace(s) to which the expected squared distance is minimized \cite{FrechetOld,karcher1977riemannian}. More specifically, a Fr\'{e}chet expectation  $\F\in\GR(n,r)$ of random subspace $\SR_k$ is a minimizer of the program
\begin{align}
 \F\in \arg \min_{\SU'\in\GR(n,r)}\mathbb{E}\l[d_{\GR}\left(\SU',\SR_k\right)^2\r],
 \label{eq:Fr\'{e}chet}
\end{align}
where the expectation is with respect to the  coefficient matrix $Q_k\in\mathbb{R}^{r\times b}$  and the support of $Y_k$.\footnote{Such a minimizer exists by Weierstrass's theorem since the objective function of Program \eqref{eq:Fr\'{e}chet} is continuous and $\GR(n,r)$ is compact. We also remark that, alternatively, one might define the Fr\'{e}chet expectation only if there exists a unique minimizer to  Program \eqref{eq:Fr\'{e}chet}.  This alternative definition will not be used here, as it does not fit the nature of our analysis.
Also, we have discarded from this expectation any matrices
for which $\operatorname{rank}(Y_k) < r$. Such matrices
can arise, with low probability, when too few samples are collected from $S_k$.
}

Is $\SR_k$ an unbiased estimator of the true subspace $\SU$? If not, how far is a Fr\'{e}chet expectation $\F$ from  $\SU$?
 We answer these questions in the rest of this section.  {Note also that $\F$ may be computed by empirically averaging the incoming sequence $\{\SR_i\}$ on the Grassmannian, as explained later in Remark~\ref{rem:imp}.}

Let us continue with a toy example with $n=2,r=1$.  For a very small $\epsilon\ll 1$, we set $S=\l[ \begin{array}{cc} \sqrt{1- \epsilon^2}, & \epsilon \end{array} \r]^*,$ so that $\SU=\mbox{span}(S)$ is nearly aligned with the first canonical vector $e_1=[1, 0]^*$. Suppose that every entry of each incoming vector is independently observed with probability of $1/2$. Therefore, every $y_t$ is  either parallel to $e_1$, or  parallel to  $e_2=[0,1]^*$,  or parallel to $S$, or  degenerate ($y_t=0$), each  with probability of $1/4$. With block size $b=1$ and after ignoring the degenerate inputs, it follows that either $\SR_k = \mbox{span}(e_1)$, or $\SR_k= \mbox{span}(e_2)$, or $\SR_k=\SU$, each with probability $1/3$. A short calculation reveals that the minimizer of Program \eqref{eq:Fr\'{e}chet}, namely the  Fr\'{e}chet expectation of $\SR_k$, is unique in this case and makes an angle of about  $\pi/6$ with $\SU$.
 That is, the Fr\'{e}chet expectation of $\SR_k$ is a biased estimator of the true subspace $\SU$, in general.

 {Perhaps this bias is somewhat unexpected, especially since each measurement block $Y_k$ satisfies $\mathbb{E}[Y_k]=pS \cdot \mathbb{E}[Q_k]$, and also because with probability one, $Q_k$ will be rank-$r$, in which case $\mbox{span}(pS\cdot Q_k) =\mbox{span}(S)=\SU$. Interestingly, results in a recent paper~\cite{chakraborty2017intrinsic} suggest that this bias disappears when averaging the span of fully observed ($p=1$) data blocks.}


 In dealing with partial samples, an important property of $\SU$ proves to be  its \emph{coherence}, defined as
\begin{equation}
\mu(\SU) : = \frac{n}{r} \max_{i\in[1:n]} \l\| S[i,:]\r\|_2^2,
\label{eq:old coh}
\end{equation}
where we use MATLAB's matrix notation to specify the rows of $S$ \cite{balzano2015local,chen2015incoherence}. One can verify that $\mu(\SU)$ is independent of  {the choice of orthobasis $S$ in \eqref{eq:old coh}}, and that $\mu\l(\SU\r)\in [1,n/r]$. For example, when $S$ consists of $r$ columns of the $n\times n$ identity matrix,  $\mu(\SU)=n/r$. In contrast, when $S$  comprises  $r$ columns of the standard Fourier matrix in $\mathbb{C}^n$,  $\mu(\SU)=1$.
Moreover, introducing  a second quantity,
\begin{equation}
\nu(\SU) := \frac{n}{r} \l\|
\l[
\begin{array}{ccc}
\l\| S[1,:]\r\|_2 & \\
& \ddots & \\
& & \l\| S[n,:]\r\|_2
\end{array}
\r]
\cdot  S^\perp \r\|^2,
\label{eq:def of coherence}
\end{equation}
will presently enable us to more tightly control the bias of the expectation of $\SR_k$. Above,  $S^\perp\in\mathbb{R}^{n\times (n-r)}$ is an orthobasis for the orthogonal complement of the subspace $\SU$. Note that $\nu(\SU)$ too is independent of the choice of orthobasis  in \eqref{eq:def of coherence} and that
\begin{equation}
\nu(\SU)\le \mu(\SU).
\label{eq:nu to mu}
\end{equation}
In the examples above, $\nu(\SU)=0$ when  $\SU$ spans $r$ columns of the identity matrix and $\nu(\SU)=1$ when $\SU$ spans $r$ columns of the Fourier matrix.  {As another example, $\nu(\SU)$ is large when $\SU=\mbox{span}(S)$ and the only nonzero entries of $S$ are $S[1,1]=S[2,1]=S[3,2]=S[4,2]=1/\sqrt{2}$.}

We are now in position to state the main result of this paper, proved in Section \ref{sec:proof of bias}. In a nutshell, this result states that the estimation bias of the  Fr\'{e}chet expectation is bounded by a factor of $\sqrt{(1\vee \frac{n}{b}) \frac{r}{pn}}$. Here and elsewhere, $a\vee b=\max(a,b)$.
\begin{thm} \textbf{\emph{(Bias of Fr\'{e}chet expectation)}}
\label{thm:bias}
Consider a subspace $\SU\in\mathbb{G}(n,r)$. Consider also a random vector $q\in\mathbb{R}^r$  and construct $Q\in\mathbb{R}^{r\times b}$ by concatenating $b$ independent copies of $q$. Let $\kappa(Q)$ be the condition number of $Q$,
and set $\Q=\mbox{span}(Q^*)\in\mathbb{G}(n,r)$. As described in Section \ref{sec:problem statement}, construct also the random subspace  $ \SR_k\in\mathbb{G}(n,r)$ and its Fr\'{e}chet expectation $\F\in\mathbb{G}(n,r)$. (The distribution of $\SR_k$ and hence its expectation are independent of $k$.)  
 Then, for any $\alpha,\widetilde{\kappa},\wt{\mu}_{\Q}\ge 1$, it holds that
\begin{align}\label{eq:bias thm}
d_{\mathbb{G}}\l(\F,\SU\r)^2
 &  \lesssim \alpha^2
\widetilde{\kappa}^2
\l(1\vee \frac{n}{b}\r) \cdot  \frac{r \l( \nu(\SU)\vee \wt{\mu}_{\Q}\r)\log(n\vee b)}{pn}
\nonumber\\
& \qquad +e^{-\alpha}+ \Pr\l[ \kappa(Q)>\widetilde{\kappa} \r] + \Pr\l[ \mu(\Q) > \wt{\mu}_{\Q}\r]
,
\end{align}
 {provided that 
\begin{align}
p \gtrsim \alpha^2 \widetilde{\kappa}^2 \l(1\vee \frac{n}{b}\r)  {\frac{r\l( \mu\l(\SU \r) \vee \wt{\mu}_{\Q} \r)\log(n\vee b)}{n}}
\label{eq:cnd on p}
\end{align}
holds.} Here, the notation $\lesssim$ suppresses any universal factors for simplicity.
\end{thm}
A few remarks are in order.
\begin{rem}\textbf{(Coefficients)}
Recall that, at time $t$, we partially observe $S\cdot q_t$, where $q_t$ is an independent copy of a  random vector $q\in\mathbb{R}^r$. The  bound in (\ref{eq:bias thm}) depends on the properties of the random matrix $Q\in\mathbb{R}^{r\times b}$, formed by concatenating $b$ independent copies of $q$.  {This dependence is indeed present though often mild in practice. Indeed, if the energy of $Q$ is concentrated, say, along its first row, then $SQ$ hardly contains any information about, say, the second column of $S$. In practice, however, it is common to assume that $q$ is a standard random Gaussian vector, in which case, $Q$ becomes a standard random Gaussian matrix.} Then, basic arguments in random matrix theory predict that
\begin{equation}
\mu\l(\Q\r) \lesssim {\log b},
\qquad \kappa(Q) \lesssim \l|\frac{\sqrt{b}+\sqrt{r}}{\sqrt{b}-\sqrt{r}}\r|,
\label{eq:dep on Q}
\end{equation}
with overwhelming probability \cite{vershynin2010introduction}.   In particular, $Q$ is well-conditioned when the block size $b$ is sufficiently large.
\end{rem}

\begin{rem}
\textbf{(Coherence)}
The bound on the bias in (\ref{eq:bias thm}) depends  {on one of the coherence factors of $\SU$, namely $\nu(\SU)$ (see (\ref{eq:def of coherence})).} This dependence suggests the estimation bias is small when $\nu(\SU)$ is small. In particular, for both of the earlier examples (column-subset of identity matrix and standard Fourier matrix), recall that $\nu$ is small.  {We point out that the role of this coherence seems inherent to the problem. Indeed, in the example with large $\nu(\SU)$ after \eqref{eq:nu to mu}, rarely are both the first and second rows observed (and, thus, nonzero).}
\end{rem}
\begin{rem}
\textbf{(Block size)} The  bound in (\ref{eq:bias thm}) also depends on the block size $b$ suggesting that, to minimize the bias, the block size should ideally be comparable to $n$, namely $b=O(n)$.  This dependence on block size was anticipated. Indeed, it is well-understood that estimating the rank-$r$ covariance   matrix of a random vector $X\in\mathbb{R}^n$ requires $O(n)$ samples in the presence of noise \cite{johnstone2012consistency}.
\end{rem}
\begin{rem}\textbf{(Measurements)}  The bound on the estimation bias in (\ref{eq:bias thm}) reduces as $p$  increases, namely as the number of measurements collected from each incoming vector increases. In particular,  $p=1$ means no erasure and $\SR_k=\SU$, if $Q$ is almost surely full-rank.
Moreover,
the bound on bias is proportional to $1/\sqrt{p}$, decreasing as $p$ increases.

\end{rem}

\begin{center}
\begin{figure}
\begin{center}
\captionsetup{justification=centering}
\includegraphics[width=0.5\textwidth]{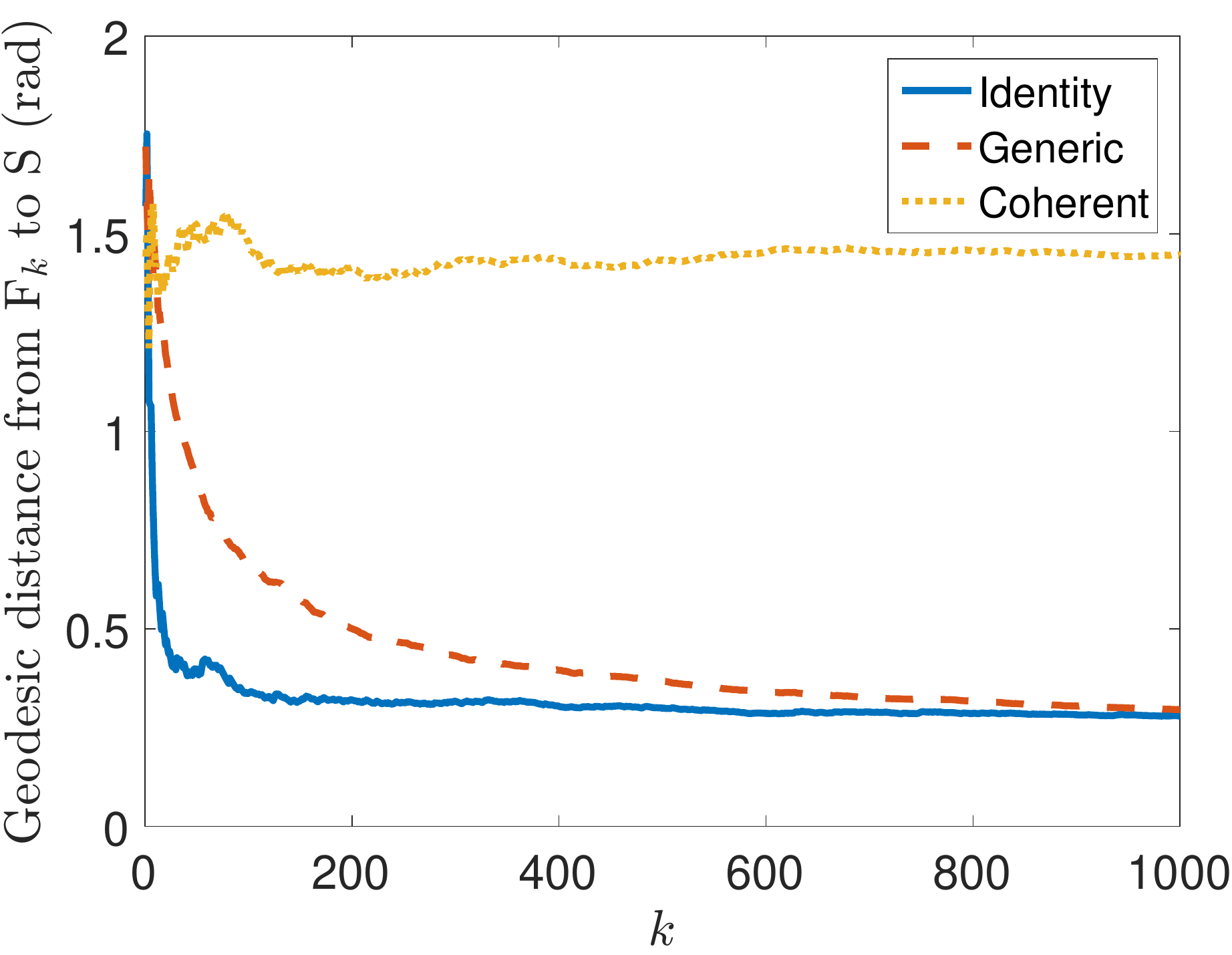}
\caption{The numerical example described in Remark \ref{rem:imp}.
\label{fig:Demo} }
\end{center}
\end{figure}
\end{center}

\begin{rem}\textbf{(Implementation)}\label{rem:imp}
 {Efficient algorithms with convergence guarantees} for computing  Fr\'{e}chet expectation  exist in the literature of computer vision and machine learning;  {the recent works  \cite{chakraborty2015recursive,salehian2015efficient,chakraborty2017intrinsic}} suit us best here. For subspaces $\A,\B\in\GR(n,r)$, consider a \emph{geodesic} connecting $\A$ and $\B$, namely a curve of shortest length connecting $\A$ and $\B$ (with respect to the canonical metric on the Grassmannian). Let $\A \#_\rho \B$ be a point on the geodesic  (itself a subspace in $\GR(n,r)$) such that $d_{\GR}(\A,\A \#_\rho \B)= \rho \cdot d_{\GR}(\A,\B)$. For example,  $\A \#_{1/2} \B$ is half way between $\A$ and $\B$. (The explicit expression for $\A \#_\rho \B$ is given in  \cite{chakraborty2015recursive}.) Suppose also, for simplicity, that  $\{\SR_k\}_k$ belongs to a geodesic ball on the Grassmannian with radius smaller than $\frac{\pi}{4}$. Then, starting with $\F_1=\SR_1$, the recursion $\F_{k}=\F_{k-1}\#_{1/k} \SR_{k}$  {converges linearly~\cite{chakraborty2017intrinsic}, in probability,} to the Fr\'{e}chet expectation $\F$, if it is unique. This recursion might be considered as a ``running average'' on the Grassmannian.

Let us consider an example with $n=50,r=2$, setting $q\in\mathbb{R}^r$ to be the standard random Gaussian vector. Entries of incoming vectors are observed with a probability of  $p=3r/n$ and measurements are partitioned into blocks of size $b=5r$. Figure \ref{fig:Demo} plots the geodesic distance $d_{\GR}(\F_k,\SU)$ versus $k$ in three cases (with $\{\F_k\}_k$ defined above): first, when $\SU$ is the span of a column subset of the identity matrix and, second, when $\SU$ is a generic subspace (say, the 	span of a standard random Gaussian matrix), and third, when $\SU$ is as described  {in the example with large $\nu(\SU)$ after \eqref{eq:nu to mu}.} In the first two cases, $\nu(\SU)$ is small (see \eqref{eq:nu to mu} and \eqref{eq:dep on Q}), predicting a relatively small estimation bias. In the third case, however, $\nu(\SU)$ and consequently the bias are large. This is indeed corroborated by Figure \ref{fig:Demo}.

\end{rem}

\section{Proof of Theorem \ref{thm:bias}}
\label{sec:proof of bias}
Let us first simplify the notation and introduce some helpful details.
For $A\in\mathbb{R}^{n\times b}$, set
\begin{equation}
\mathcal{P}_p(A) := \sum_{i=1}^n \sum_{j=1}^b \epsilon_{i,j} A[i,j] \cdot E_{i,j},
\label{eq:def of R}
\end{equation}
where $\{\epsilon_{i,j}\}_{i,j}$ is a sequence of independent Bernoulli random variables, taking one with probability $p$ and zero otherwise. Also, $E_{i,j}\in\mathbb{R}^{n\times b}$ is the $[i,j]$-th canonical matrix in $\mathbb{R}^{n\times b}$, i.e.,   $E_{i,j}[i,j]=1$ is the only nonzero entry of $E_{i,j}$.   Let $\Omega\subset [1:n]\times [1:b]$ be the random index set corresponding to the support of $\mathcal{P}_p(A)$.
We set $Y=\mathcal{P}_p(SQ)$ and let $Y_r\in\mathbb{R}^{n\times b}$ be a rank-$r$ truncation of $Y$, obtained via singular value decomposition (SVD). We also  let $\SR=\mbox{span}(Y_r)$. Note that $\SR$ is a random subspace on  {the Grassmannian $\mathbb{G}(n,r)$.}

We wish to calculate how far the true subspace $\mbox{S}$ is from  Fr\'{e}chet expectation(s)  of $\SR$, defined as solution(s) of the program
\begin{align}
\min_{\SU'\in\GR(n,r)} f\l( \SU'\r),
\qquad
f\l( \SU'\r):= \mathbb{E}\Big[d_{\GR}\left(\SU',\SR\right)^{2}\Big],
 \label{eq:re def Fr\'{e}chet}
\end{align}
where the expectation is with respect to the coefficient matrix $Q$ and the support  $\Omega$.
Let $\{\theta_i(\SU',\SR)\}_{i=1}^r$ denote the principal angles between the two subspaces $\SU',\SR\in\GR(n,r)$. It is well-known that  $\{\sin(\theta_i(\SU',\SR))\}_{i=1}^r$ are in fact the singular values of $P_{\SU'^\perp}P_{\SR}$,  {where $P_{\operatorname{A}}$ denotes the orthogonal projection matrix onto a subspace $\operatorname{A}$.} Moreover,
\begin{align}
d_{\GR}\l( \SU',\SR\r)^2  = \frac{1}{r} \sum_{i=1}^r \theta_i\l(\SU',\SR \r)^2
 =
\frac{1}{r} \l\| \mbox{arcsin}\l(P_{\SU'^\perp}P_{\SR} \r) \r\|_F^2,
\label{eq:exp for dGR}
\end{align}
where $\mbox{arcsin}(\cdot)$, applied to a matrix, acts only on the singular values, leaving the singular vectors intact  \cite{golub2013matrix}.
 The geodesic distance $d_{\GR}(\SU',\SR)$ is tightly controlled as follows:
\begin{align}
 d_{\GR}\l(\SU',\SR\r)^2 
& = \frac{1}{r} \l\| \mbox{arcsin}\l(P_{\SU'^\perp}P_{\SR} \r)\r\|_F^2 
\qquad \mbox{(see \eqref{eq:exp for dGR})}
\nonumber\\
& \lesssim \frac{1}{r} \l\|  P_{\SU'^\perp}P_{\SR} \r\|_F^2 
\qquad \l( |\arcsin a| \le {\pi}|a|/2\r)
\nonumber\\
& \le  \l\|  P_{\SU'^\perp}P_{\SR} \r\|^2,
\quad \l( \mbox{rank}\l(P_{\SU'^\perp}P_{\SR} \r) \le r\r) \label{eq:bnd on d}
\end{align} 
\begin{align}
d_{\GR}\l(\SU',\SR \r)^2 
& = \frac{1}{r} \sum_{i=1}^r  \theta_i\l(\SU',\SR \r)  ^2 
\qquad \mbox{(see \eqref{eq:exp for dGR})}
\nonumber\\
& \ge \frac{1}{r} \sum_{i=1}^r  \sin^2\l( \theta_i\l(\SU',\SR \r)   \r) 
\qquad \l(|a|\ge |\sin a|\r) \nonumber\\
& = \frac{1}{r}\l\| P_{\SU'^\perp}P_{\SR} \r\|_F^2.
\label{eq:lwr bnd on d}
\end{align}
In turn, \eqref{eq:bnd on d} and \eqref{eq:lwr bnd on d} allow us to tightly control $f(\SU')$ for arbitrary $\SU'\in\GR(n,r)$:
\begin{align}
f\l(\SU'\r) 
& = \frac{1}{r}\mathbb{E}\l[ d_{\GR}\l(\SU',Y \r)^2\r]
\qquad \mbox{(see \eqref{eq:re def Fr\'{e}chet})}\nonumber\\
& \lesssim  \E \l[ \l\| P_{\SU'^\perp } P_{\SR}\r\|^2 \r], 
\qquad \mbox{(see \eqref{eq:bnd on d})}
\label{eq:upp bnd on f}
\end{align}
\begin{align}
f\l( \SU' \r) 
& \ge \frac{1}{r}\E \l[\l\| P_{\SU'^\perp}P_{\SR} \r\|_F^2 \r].
\qquad \mbox{(see \eqref{eq:lwr bnd on d})}
\label{eq:lwr bnd on f}
\end{align}
Of particular interest to us is evaluating $f(\SU)$ which, as \eqref{eq:upp bnd on f} and \eqref{eq:lwr bnd on f}  suggest,  requires an estimate of the principal angle $\sin(\theta_1(\SU,\SR))=\|P_{\SU^\perp}P_{\SR}\|$. The following result is proved in Section \ref{sec:Proof of Lemma angle S n R} with the aid of  a sharp perturbation bound, as well as basic large deviation results. 
\begin{lem}\label{lem:angle S n R}
Fix $Q\in\mathbb{R}^{r\times b}$ with rank $r$ and let $\kappa(Q)$ be its condition number. Also set $\Q = \mbox{span}(Q^*)$. 
For $\alpha\ge1$ and except with a probability of at most $e^{-\alpha}$, it holds that 
\begin{align*}
& \l\| P_{\SU^\perp} P_{\SR}\r\| \nonumber\\
 & \lesssim \alpha \cdot \kappa(Q) \l( 1 \vee \sqrt{\frac{n}{b}} \r)   \sqrt{\frac{r \l(\nu(\SU)\vee \mu(\Q) \r) \log\l(n\vee b \r)}{pn}},
\end{align*}
provided that 
\begin{align*}
 p \gtrsim \alpha^2 \kappa(Q)^2 \l(1\vee \frac{n}{b}\r)  {\frac{r\l( \mu\l(\SU \r) \vee \mu(\Q) \r)\log(n\vee b)}{n}}.
\end{align*}
\end{lem}
 Lemma \ref{lem:angle S n R} readily translates into a bound on $f(\SU)$:  
For $\alpha,\widetilde{\kappa},\wt{\mu}_{\Q}\ge 1$, suppose that 
\begin{align}
p \gtrsim \alpha^2 \widetilde{\kappa}^2 \l(1\vee \frac{n}{b}\r)  {\frac{r\l( \mu\l(\SU \r) \vee \wt{\mu}_{\Q} \r)\log(n\vee b)}{n}},
\label{eq:cnd on p}
\end{align}
and conveniently  
 let $\mathcal{E}$ denote the event where both $\kappa(Q)\le \wt{\kappa}_{\Q}$ and $\mu(Q)\le \wt{\mu}_{\Q}$ hold. Also set 
 \begin{equation*}
\Delta := \alpha \cdot \kappa(Q) \l(1\vee \sqrt{\frac{n}{b}}  \r) \sqrt{\frac{r \l(\nu\l(\SU\r) \vee \mu\l(\operatorname{Q} \r)  \r) \log\l( n\vee b\r)}{pn}}.
\end{equation*}
and let $\mathcal{E}'$ be the event where $\l\|P_{\SU}^{\perp}P_{\SR}\r\|^2  \le \Delta^2$. Then, note that
\begin{align}
f\l( \SU \r) 
& \lesssim  \E\l[ \l\| P_{\SU^\perp} P_{\SR} \r\|^2 \r] 
\qquad \mbox{(see \eqref{eq:upp bnd on f})}
\nonumber\\
& =  \E\l[ \l\| P_{\SU^\perp} P_{\SR} \r\|^2 | \mathcal{E}\r] \cdot \Pr\l[\mathcal{E}\r] \nonumber\\
& \qquad + \E\l[ \l\| P_{\SU^\perp} P_{\SR} \r\|^2 | \mathcal{E}^C\r] \cdot \Pr\l[\mathcal{E}^C\r] \nonumber\\
& \le   \E\l[ \l\| P_{\SU^\perp} P_{\SR} \r\|^2 | \mathcal{E}\r] 
+  \Pr\l[\mathcal{E}^C \r],
\label{eq:added recently}
\end{align}
where $\mathcal{E}^C$ is the complement of the event $\mathcal{E}$. To bound the expectation in the last line above, let $\mathcal{E}'$ be the event where $\l\|P_{\SU}^{\perp}P_{\SR}\r\|^2  \le \Delta^2$ and note that 
\begin{align}
& \E\l[ \l\| P_{\SU^\perp} P_{\SR} \r\|^2 | \mathcal{E}\r] \nonumber\\
& =  \E_{Q} \l[ \E_{\Omega}  
\l[ \l\| P_{\SU^\perp} P_{\SR} \r\|^2 \r] | \mathcal{E} \r] \nonumber\\
& = \mathbb{E}_Q\l[ \mathbb{E}_\Omega \l[ 
\l\|P_{\SU}^{\perp}P_{\SR}\r\|^2 
| \mathcal{E}'  \r] \Pr\l[\mathcal{E}' \r]
\r.
\nonumber\\
&\qquad  \l.+ \mathbb{E}_{\Omega}\l[
\l\|P_{\SU}^{\perp}P_{\SR}\r\|^2 
| \mathcal{E}'^C  \r] \Pr\l[\mathcal{E}'^C \r]
| \mathcal{E} \r] 
\nonumber\\
& \le \mathbb{E}_Q\l[  \Delta^2 +e^{-\alpha} |\mathcal{E}\r]
\qquad \mbox{(see Lemma \ref{lem:angle S n R})}
\nonumber\\
& \le \Delta^2 +e^{-\alpha}
\label{eq:added recently 2}
\end{align}
Substituting the bound above back into \eqref{eq:added recently}, we find that 
\begin{align}
& f(\SU) \nonumber\\
& \le     \E\l[ \l\| P_{\SU^\perp} P_{\SR} \r\|^2 | \mathcal{E}\r]
+  \Pr\l[\mathcal{E}^C \r]
\qquad \mbox{(see \eqref{eq:added recently})} \nonumber\\
& \lesssim  \Delta^2 + e^{-\alpha}+ \Pr\l[\mathcal{E}^C \r].
\qquad \mbox{(see \eqref{eq:added recently 2})}
\label{eq:final upp bnd on fS}
\end{align}
In words, when $p$ is sufficiently large, $f(\SU)$ is small. 
Let us next find a lower bound on $f(\cdot)$ far from $\SU$:  For an arbitrary  subspace $\mbox{S}'\in\GR(n,r)$, we note that 
\begin{align}
& f\left(\mbox{S}'\right) \nonumber\\
& \ge \frac{1}{r}\E\l[ \l\| P_{\SU'^\perp} P_{\SR}  \r\|_F^2 \r] 
\qquad \mbox{(see \eqref{eq:lwr bnd on f})}
\nonumber\\
& \ge \frac{1}{r}\l\| P_{\SU'^\perp} \cdot  \E\l[ P_{\SR} \r]   \r\|_F^2 
\qquad \mbox{(Jensen's ineq.)}
\nonumber\\
 & \ge \frac{1}{r}\left(\left\Vert P_{\SU'^{\perp}}P_{\SU}\right\Vert _{F}-\left\Vert P_{\SU'^{\perp}}\left(P_{\SU}-\mathbb{E}\left[P_{\SR}\right]\right)\right\Vert _{F}\right)^{2}\,
\mbox{(triangle ineq.)} 
 \nonumber \\
 & \ge \frac{1}{2r}\left\Vert P_{\SU'^{\perp}}P_{\SU}\right\Vert _{F}^2- \frac{1}{r}\left\Vert P_{\SU'^{\perp}}\left(P_{\SU}-\mathbb{E}\left[P_{\SR}\right]\right)\right\Vert _{F}^2
 \nonumber\\
 & \ge\frac{1}{2r}\left\Vert P_{\SU'^{\perp}}P_{\SU}\right\Vert _{F}^{2}-\frac{1}{r}\left\Vert P_{\SU}-\mathbb{E}\left[P_{\SR}\right]\right\Vert _{F}^{2}
 \nonumber \\
 & \ge \frac{1}{2} \l\| P_{\SU'^\perp}P_{\SU} \r\|_F^2- \E\l[ \left\Vert P_{\SU}- P_{\SR} \right\Vert _{F}^{2} \r]
\qquad \mbox{(Jensen's ineq.)}
 \nonumber \\
 & = \frac{1}{2r} \l\| P_{\SU'^\perp} P_{\SU} \r\|_F^2- \frac{2}{r} \E\l[ \left\Vert P_{\SU^\perp}P_{\SR} \right\Vert _{F}^{2} \r]
 \nonumber\\
 & \ge \frac{1}{2r} \l\|P_{\SU'^\perp} P_{\SU}  \r\|_F^2
 -2  \mathbb{E}\left\Vert P_{\SU^\perp}P_{\SR}\right\Vert^{2}\qquad
\l( \mbox{rank} \l( P_{\SU}^\perp P_{\SR} \r) \le r  \r) 
 \nonumber \\
&  \ge \frac{1}{2r} \l\|P_{\SU'^\perp} P_{\SU}  \r\|_F^2
 -2 \Cl{global} \l( \Delta^2 + e^{-\alpha}+\Pr\l[ \mathcal{E}^C \r] \r),
 \,  \mbox{(see \eqref{eq:final upp bnd on fS})}\label{eq:lowr bnd on f}
\end{align}
for an absolute constant $\Cr{global}>0$. Above, we used the inequality $\l( a-b\r)^2 \ge \frac{a^2}{2}-b^2$ for scalars $a,b$ and the fact that $\l\| P_{\SU} - P_{\SR} \r\|_F^2 = 2 \l\| P_{\SU^\perp} P_{\SR} \r\|_F^2$. 
To summarize in words, $f(\SU)$ is small for large $p$  because of (\ref{eq:final upp bnd on fS}). Moreover, thanks to (\ref{eq:lowr bnd on f}), we know that $f(\mbox{S}')$ is large for any subspace $\SU'$ far from $\SU$. Therefore, any minimizer of $f(\cdot)$ in \eqref{eq:re def Fr\'{e}chet} (namely, any Fr\'{e}chet expectation $\F$) must be close
to the true subspace $\mbox{S}$. More formally,
\begin{equation} 
\l\| P_{\SU'^\perp}P_{\SU} \r\|_F^2
\gtrsim  \Delta^2 + \Pr\l[ \mathcal{E}^C \r] 
\Longrightarrow
f\l( \SU'\r) > f\l( \SU\r),
\end{equation}
and, consequently, 
\begin{equation}
\frac{1}{r}{\sum_{i=1}^r \sin^2\l( \theta_i\l( \F,\SU \r) \r)} = \frac{\l\| P_{\F^\perp} P_{\SU}\r\|_F^2}{r} \lesssim  \Delta^2 + e^{-\alpha}+ \Pr\l[ \mathcal{E}^C \r], 
\end{equation}
which completes the proof of Theorem \ref{thm:bias} after noting that $\Pr[\mathcal{E}^C]\le \Pr[\kappa(Q)>\wt{\kappa}]+ \Pr[\mu(\Q)>\wt{\mu}_{\Q}]$, and using the fact that $|a|\le \pi |\sin(a)|/2$ when $|a|\le \pi /2$.

\bibliographystyle{unsrt}
\bibliography{References}

\section{Proof of Lemma \ref{lem:angle S n R}}
\label{sec:Proof of Lemma angle S n R}
Fix $Q\in\mathbb{R}^{r\times  b}$ for now and assume $Q$ is rank-$r$. Consider the measurement matrix $Y=\mathcal{P}_p(SQ)\in\mathbb{R}^{n\times b}$ and let $Y_r\in\mathbb{R}^{n\times b}$ be a rank-$r$ truncation of $Y$, obtained via SVD, and set $\SR=\mbox{span}(Y_r)$. Let also $Y_{r^+}:=Y-Y_r$ denote the residual. Note that 
\begin{align}
 \l\| P_{\SU^\perp}P_{\SR} \r\|  
& = \l\| P_{\SU^\perp} Y_rY_r^{\dagger} \r\| \nonumber\\
& = \l\| P_{\SU^\perp} \l(Y-Y_{r^+} \r) Y_r^{\dagger} \r\| \nonumber\\
& = \l\| P_{\SU^\perp} Y Y_r^{\dagger} \r\|  
\qquad \l( Y_{r^+} Y_r^*=0  \r) \nonumber\\
& \le \l\| P_{\SU^\perp} Y\r\| \cdot \l\| Y_r^{\dagger}\r\| \nonumber\\
& = \frac{\l\| P_{\SU^\perp} Y \r\| }{\sigma_r\l(Y_r \r)} \nonumber\\
& \le \frac{\l\| P_{\SU^\perp} Y \r\| }{\sigma_r\l(pSQ \r)- \l\| Y_r-pSQ \r\|}
\,\,\,\,\,\,\,\, \mbox{(Weyl's inequality)}
 \nonumber\\
& = \frac{\l\| P_{\SU^\perp} Y \r\| }{p\cdot \sigma_r\l(Q \r)- \l\| Y_r-pSQ \r\|},
\qquad \l(S^*S=I_r \r)
\label{eq:tail bound pert}
\end{align}
which is slightly sharper than the standard perturbation bound \cite[Theorem 3]{wedin1972perturbation}, and the difference is consequential in our problem. 
We next control both norms in the last line above. Beginning with the numerator, we write that 
\begin{align}
P_{\SU^\perp} Y & = P_{\SU^\perp} \mathcal{P}_p\l(SQ\r) \nonumber\\
& = P_{\SU^\perp} \sum_{i,j} \epsilon_{i,j} \cdot (SQ)[i,j] \cdot E_{i,j}  
\qquad \mbox{(see \eqref{eq:def of R})} \nonumber\\
& =: \sum_{i,j} Z_{i,j},
\label{eq:def Zs}
\end{align}
where $\{Z_{i,j}\}_{i,j}\subset \mathbb{R}^{n\times b}$ are independent zero-mean random matrices. In order to appeal to the matrix Bernstein
inequality \cite{tropp2012user}, some preparation is required: 
\begin{align}
& \nu(\SU) \nonumber\\
& = \frac{n}{r}\l\|
\l[
\begin{array}{ccc}
\l\| S[1,:]\r\|_2 & \\
& \ddots & \\
& & \l\| S[n,:]\r\|_2
\end{array}
\r]
\cdot  S^\perp \r\|^2 
\qquad \mbox{(see \eqref{eq:def of coherence})}
\nonumber\\
& = \frac{n}{r} \l\|\sum_{i=1}^n \l\| S[i,:] \r\|_2^2 \cdot P_{\SU^\perp} E_{i,i} P_{\SU^\perp} \r\| \nonumber\\
& \le e \frac{n}{r} \max_i \l\| \l\| S[i,:] \r\|_2^2 \cdot P_{\SU^\perp} E_{i,i} P_{\SU^\perp} \r\| \nonumber\\
& = \frac{n}{r} \max_i \l(  \l\| S[i,:] \r\|_2^2 \cdot \l\| S^\perp[i,:]\r\|_2^2 \r),
\label{eq:new coh simplified}
\end{align}
\begin{align}
& \max_{i,j}\l\|Z_{i,j} \r\| \nonumber\\
& = \max_{i,j} \l\| \epsilon_{i,j} \cdot (SQ)[i,j]\cdot P_{\SU^\perp}E_{i,j}\r\|
\qquad \mbox{(see \eqref{eq:def Zs})} \nonumber\\
& \le \max_{i,j} \l\|  (SQ)[i,j]\cdot P_{\SU^\perp}E_{i,j}\r\|
\qquad \l( \epsilon_{i,j}\in\{0,1\} \r) \nonumber\\
& \le  \max_{i,j} \l\|S[i,:] \r\|_2 \cdot  \l\| Q[:,j] \r\|_2 \cdot \l\|   P_{\SU^\perp} E_{i,j} \r\| \nonumber\\
& \le  \|Q\|\sqrt{\frac{r\mu(\Q)}{b}} \max_{i,j} \l\|S[i,:] \r\|_2 \cdot \l\|  P_{\SU^\perp} E_{i,j} \r\|
\qquad \mbox{(see \eqref{eq:old coh})} \nonumber\\
& =  \|Q\|\sqrt{\frac{r\mu(\Q)}{b}} \max_{i}  \l\|S[i,:] \r\|_2 \cdot\l\| S^\perp[i,:] \r\|_2 
 \nonumber\\
& \le \|Q\|\sqrt{\frac{r\mu(\Q)}{b}} \sqrt{\frac{r\nu(\SU)}{n}} =: \beta,
\qquad \mbox{(see \eqref{eq:new coh simplified})}
\label{eq:beta} 
\end{align}
\begin{align}
&\l\| \E_\Omega\l[\sum_{i,j} Z_{i,j}Z_{i,j}^* \r]\r\| \nonumber\\
& = \l\|  \sum_{i,j} \E_\Omega \l[\epsilon_{i,j}^2 \r]  \cdot \l|(SQ)[i,j]  \r|^2 \cdot P_{\SU^\perp} E_{i,i}P_{\SU^\perp} \r\|
\nonumber\\
& = \l\|  \sum_{i,j} \E_\Omega \l[\epsilon_{i,j} \r]  \cdot \l|(SQ)[i,j]  \r|^2 \cdot P_{\SU^\perp} E_{i,i}P_{\SU^\perp} \r\|
\,\,\l( \epsilon_{i,j}\in\{0,1\} \r) \nonumber\\
& = p \l\|  \sum_{i,j}   \l|(SQ)[i,j]  \r|^2 \cdot P_{\SU^\perp} E_{i,i}P_{\SU^\perp} \r\|
\,\,\, \l(\epsilon_{i,j}\sim \mbox{Bernoulli}(p)  \r) \nonumber\\
& =p \l\|  \sum_{i}   \l\|S[i,:]Q  \r\|_2^2  \cdot P_{\SU^\perp} E_{i,i}P_{\SU^\perp} \r\| \nonumber\\
& \le p\|Q\|^2 \l\|  \sum_{i}   \l\|S[i,:]  \r\|_2^2  \cdot P_{\SU^\perp} E_{i,i}P_{\SU^\perp} \r\| \nonumber\\
& = 
p\|Q\|^2 \frac{r\nu(\SU)}{n},
\qquad \mbox{(see \eqref{eq:new coh simplified})} \label{eq:sigma leg 1}
\end{align}
\begin{align}
&\l\| \E_\Omega\l[\sum_{i,j} Z_{i,j}^*Z_{i,j} \r]\r\| \nonumber\\
& = \l\|  \sum_{i,j} \E_\Omega \l[\epsilon_{i,j} \r]  \cdot \l|(SQ)[i,j]  \r|^2 \cdot  E_{j,i}P_{\SU^\perp}E_{i,j} \r\| \nonumber\\
& = p \l\|  \sum_{i,j} \l|(SQ)[i,j]  \r|^2 \cdot  E_{j,i}P_{\SU^\perp}E_{i,j} \r\| 
\,\,\,\, \l(  \epsilon_{i,j}\sim \mbox{Bernoulli}(p) \r) 
\nonumber\\
& \le p \l\|  \sum_{i,j} \l|(SQ)[i,j]  \r|^2 \cdot  E_{j,i}E_{i,j} \r\|  
\nonumber\\
& = p \l\|  \sum_{i,j} \l|(SQ)[i,j]  \r|^2 \cdot  E_{j,j} \r\|  
\nonumber\\
& = p \l\| \sum_{j} \l\| S\cdot Q[:,j] \r\|_2^2  \cdot E_{j,j}\r\|  \nonumber\\
& = p \l\| \sum_{j} \l\|  Q[:,j] \r\|_2^2  \cdot E_{j,j}\r\| 
\qquad \l( S^* S = I_r \r) \nonumber\\
& = p  \max_j \l\| Q[:,j] \r\|_2^2 \nonumber\\
& \le p \|Q\|^2  \frac{r\mu(\Q)}{b},
\qquad \mbox{(see \eqref{eq:old coh})}
\label{eq:sigma leg 2}
\end{align}
\begin{align}
& \sigma^2 \nonumber\\
& :=  \l\|\mathbb{E}_\Omega\l[\sum_{i,j} Z_{i,j}Z_{i,j}^*  \r] \r\| 
\vee \l\|\mathbb{E}_\Omega\l[\sum_{i,j} Z_{i,j}^* Z_{i,j}  \r] \r\|
 \nonumber\\
& \le p \|Q\|^2 \l( 1 \vee \frac{n}{b} \r)\cdot \frac{r \l(\nu(\SU)\vee \mu(\Q)\r)}{n} .
\quad \,\,\,\,\mbox{(see \eqref{eq:sigma leg 1} and \eqref{eq:sigma leg 2})}
\label{eq:sigma def}
\end{align}
(In fact, using a slightly different argument, $p$ in \eqref{eq:sigma def} can be replaced with $p(1-p)$. However, since $p$ is typically small, this does not lead to a substantial improvement in final results and is therefore  ignored here.) We are now in position to apply the matrix Bernstein inequality \cite{tropp2012user}: For $\alpha\ge 1$ and except with a probability of at most $e^{-\alpha}$, it holds that 
\begin{align}
&    \left\Vert P_{\SU^\perp} Y\right\Vert \nonumber\\
& = \l\| \sum_{i,j}Z_{i,j} \r\|
\qquad \mbox{(see \eqref{eq:def Zs})}
 \nonumber\\
    & \lesssim \alpha \cdot  \max\l( \log(n\vee b) \cdot  \beta, \sqrt{\log (n\vee b)}\cdot \sigma\r) 
   \nonumber\\
& \le    \alpha\|Q\| \l(1 \vee \sqrt{\frac{n}{b}}  \r)\cdot\max\left(
\log(n\vee b ) \cdot \frac{r \l(\nu(\SU)\vee \mu(\Q)  \r)}{n}
\r. \nonumber\\
& \qquad \qquad \l. ,\sqrt{\log (n\vee b)} \cdot \sqrt{p} \cdot \sqrt{\frac{r\l( \nu(\SU)\vee \mu(\Q)\r)}{n}} \r) 
\nonumber\\
& = \alpha \|Q\| \l(1 \vee \sqrt{\frac{n}{b}} \r) \cdot \sqrt{\log(n\vee b)} \cdot \sqrt{p} \cdot \sqrt{\frac{r\l( \nu(\SU)\vee \mu(\Q)\r)}{n}},
\label{eq:Bernie}
\end{align}
when $p$ is sufficiently large. On the other hand, note that $\E_\Omega[\mathcal{P}_p(SQ)]=pSQ$. In fact, another application of the matrix Bernstein's inequality proves that 
\begin{align}
& \left\Vert  \mathcal{P}_{p}(SQ) - pSQ\right\Vert \nonumber\\
& \le C \alpha \|Q\| \l(1 \vee \sqrt{\frac{n}{b}} \r) \cdot \sqrt{\log(n\vee b)} \cdot 
\sqrt{\frac{pr\l( \mu\l(\SU \r) \vee \mu(\Q)\r)}{n}},
\end{align}
for some constant $C$. 

In particular, with $\kappa(Q)=\|Q\|/\sigma_r(Q)$ denoting the condition number of $Q$ and after taking
\begin{equation}
 p \ge C^2 \alpha^2 \kappa(Q)^2 \l(1\vee \frac{n}{b}\r)  {\frac{r\l( \mu\l(\SU \r) \vee \nu(\Q) \r)\log(n\vee b)}{n}},
 \label{eq:p large enough}
\end{equation}
we find that 
\begin{align}
\left\Vert \mathcal{P}_{p}(SQ) -pSQ\right\Vert \le \frac{p\cdot \sigma_r(Q)}{2}.
 \label{eq:small enough}
\end{align}
 Plugging \eqref{eq:Bernie} and \eqref{eq:small enough} back into (\ref{eq:tail bound pert})
yields that 
\begin{align*}
& \l\| P_{\SU^\perp} P_{\SR}\r\| \nonumber\\ 
& \le 
\frac{\l\| P_{\SU^\perp} Y \r\| }{p\cdot \sigma_r\l(Q \r)- \l\| Y_r-pSQ \r\|}
\qquad \mbox{(see \eqref{eq:tail bound pert})} \nonumber\\
& \le \frac{\l\| P_{\SU^\perp} Y \r\| }{p\cdot \sigma_{r}(Q)/2}
\qquad \mbox{(see \ref{eq:small enough})} \nonumber\\
& \lesssim \alpha \cdot \kappa(Q) \l( 1 \vee \sqrt{\frac{n}{b}} \r) \cdot  \sqrt{\frac{r \l( \nu(\SU) \vee \mu(\Q) \r) \log\l(n\vee b \r)}{pn}} ,
\end{align*}
provided that $p$ satisfies \eqref{eq:p large enough}. The last line above uses \eqref{eq:Bernie}. This completes the proof of Lemma \ref{lem:angle S n R}.

\end{document}